\documentclass{article}

\usepackage{arxiv}
\usepackage{setspace}

\usepackage[utf8]{inputenc} 
\usepackage[T1]{fontenc}    
\usepackage{hyperref}       
\usepackage{url}            
\usepackage{booktabs}       
\usepackage{amsfonts}       
\usepackage[numbers]{natbib}
\usepackage{amsmath, enumitem}
\usepackage{upgreek}
\usepackage{graphicx}
\graphicspath{{Figures/}} 
\usepackage{setspace}
\usepackage{multirow}
\usepackage{fixltx2e}
\usepackage{float}

\title{Solving All Seismic Tomographic Problems using Deep Learning}

\date{} 					

\author{
  Xin Zhang \\
  State Key Laboratory of Deep Earth Exploration and Imaging \\
  School of Engineering and Technology \\
  China University of Geosciences\\
  Beijing, China \\
  \texttt{xzhang@cugb.edu.cn} \\
   \And
  Kaiwen Xia \\
  State Key Laboratory of Deep Earth Exploration and Imaging \\
  School of Engineering and Technology \\
  China University of Geosciences\\
  Beijing, China \\
}

\begin{document}
\maketitle

\begin{abstract}
In a variety of geoscientific applications scientists often need to image properties of the Earth's interior in order to understand the heterogeneity and processes taking place within the Earth. Seismic tomography is one such method which has been used widely to study properties of the subsurface. In order to solve tomographic problems efficiently, neural network-based methods have been introduced to geophysics. However, these methods can only be applied to certain types of problems with fixed acquisition geometry at a specific site. In this study we extend neural network-based methods to problems with various scales and acquisition geometries by using graph mixture density networks (MDNs). We train a graph MDN for 2D tomographic problems using simulated velocity models and travel time data, and apply the trained network to both synthetic and real data problems that have various scales and station distributions at different sites. The results demonstrate that graph MDNs can provide comparable solutions to those obtained using traditional Bayesian methods in seconds, and therefore provide the possibility to use graph MDNs to produce rapid solutions for all kinds of seismic tomographic problems over the world.
\end{abstract}

\section{Introduction}
Seismic tomography is a methodology to infer properties of the Earth's interior using seismic data recorded at the surface. The subsurface is usually parameterized in some way, and seismic tomography solves a parameter estimation problem. Tomographic problems are traditionally solved using optimization methods by minimizing the difference between observed data and model predicted data, together with some form of regularization \cite{aki1976determination, aster2018parameter}. However, these methods cannot provide accurate estimates of uncertainty because they assume a linearized physics. Bayesian inference methods have therefore been introduced to seismic tomography to quantify uncertainty, including Monte Carlo \cite{mosegaard1995monte, tarantola2005inverse} and variational inference \cite{zhang2020seismic, zhang20233}. However, these methods generally require expensive computational cost, and cannot be applied in many applications that need efficient solutions. In addition, both optimization and Bayesian methods require tedious and bespoke tuning of hyperparameters, and therefore cannot be performed automatically and efficiently for practical applications.

Neural network-based methods provide an automatic and efficient alternative to solve tomographic problems. In these methods one trains a neural network to emulate the inverse mapping from data to parameter space using a set of example pairs of data and model parameters \cite{bishop2006pattern}. Once trained, the network can provide rapid estimates of model parameters for any new datasets. These methods have been applied to a variety of geophysical inverse problems \cite{roth1994neural, araya2018deep, yu2021deep}. In addition, probabilistic forms of neural networks have been introduced to provide estimates of Bayesian posterior distributions in a range of applications \cite{devilee1999efficient, meier2007aglobal, earp2020probabilistic, zhang2021bayesianb}. However, in these studies the networks are trained for a specific acquisition geometry and a fixed number of data, and consequently can only be applied to similar problems with new datasets. To reduce this issue, we introduced graph mixture density networks (MDNs) to solve seismic tomographic problems by representing seismic travel time data using a graph, and demonstrated that the method can be applied to problems with variable sizes of data for a given acquisition geometry \cite{zhang2024rapid}.

In this study we extend graph MDNs to solve seismic tomographic problems with variable scales and acquisition geometries. To achieve this, we train the neural network using travel time data simulated from random velocity structures and acquisition geometries, so that the trained network can be directly applied to problems with variable source and receiver distributions. In addition, since the travel time data, the acquisition geometry and the velocity structure can be scaled proportionally across different ranges, the network can theoretically be applied to different scales. We demonstrate the method by training a graph MDN in a small region, and apply the trained network to both synthetic and real data examples with larger sizes. The results show that the network can produce accurate approximation of posterior distributions to those obtained using Monte Carlo and variational inference methods in seconds for problems with various scales and acquisition geometries. This therefore demonstrates the possibility to train one or a set of deep learning models to solve all kinds of seismic tomographic problems over the world.      

\section{Methods}
\subsection{Bayesian inference}
Bayesian inference constructs a posterior probability density function (pdf) $p(\mathbf{m}|\mathbf{d}_{\mathrm{obs}})$ of model parameters $\mathbf{m}$ given observed data $\mathbf{d}_{\mathrm{obs}}$ by updating a prior pdf $p(\mathbf{m})$ with new information contained in the data. According to Bayes' theorem,
\begin{equation}
	p(\mathbf{m}|\mathbf{d}_{\mathrm{obs}})=\frac{p(\mathbf{d}_{\mathrm{obs}}|\mathbf{m})p(\mathbf{m})}{p(\mathbf{d}_{\mathrm{obs}})}
	\label{eq:Bayes}
\end{equation}
where $p(\mathbf{d}_{\mathrm{obs}}|\mathbf{m})$ is the likelihood function which describes the probability of observing data $\mathbf{d}_{\mathrm{obs}}$ if the model $\mathbf{m}$ was true, and $p(\mathbf{d}_{\mathrm{obs}})$ is a normalization factor called the evidence. The likelihood function is often assumed to be a Gaussian distribution around the data predicted synthetically from model $\mathbf{m}$.

\subsection{Mixture density networks}
Mixture density networks (MDNs) are a class of neural networks that output a probability density function for any input. This is achieved by representing the pdf using a mixture model, which is defined as a linear combination of kernel functions, and can be used to provide approximation to an arbitrary probability distribution $p(\mathbf{m}|\mathbf{d})$,
\begin{equation}
	p(\mathbf{m}|\mathbf{d}) \approx \sum_{i=1}^{N}\alpha_{i}(\mathbf{d})\phi_{i}(\mathbf{m}|\mathbf{d})
	\label{eq:mixture_model}
\end{equation}
where $\phi_{i}(\mathbf{m}|\mathbf{d})$ is the $i^{th}$ kernel function, $\alpha_{i}$ is the weight (called mixture coefficient) of the $i^{th}$ kernel such that $\alpha_{i}>0$ and $\sum_{i=1}^{N}\alpha_{i}=1$, and $N$ is the number of components in the mixture \cite{mclachlan1988mixture}. In this study we use a Gaussian kernel function,
\begin{equation}
	\phi_{i}(\mathbf{m}|\mathbf{d}) = \frac{1}{\prod_{k=1}^{c}\left( \sqrt{2\pi}\sigma_{ik}(\mathbf{d}) \right)}
	\mathrm{exp}\left\{ -\frac{1}{2} \sum_{k=1}^{c} \frac{\left(m_{k} - \mu_{ik}(\mathbf{d}) \right)^{2}}{\sigma_{ik}^{2}(\mathbf{d})} \right\}
	\label{eq:gaussian_kernel}
\end{equation}
where $\mu_{ik}$ and $\sigma_{ik}$ are the $k^{th}$ elements of the mean and standard deviation of the $i^{th}$ Gaussian kernel, and $c$ is the number of elements in $\mathbf{m}$. Note that although equation \ref{eq:gaussian_kernel} assumes a diagonal covariance matrix for each Gaussian distribution, the mixture model in equation \ref{eq:mixture_model} is sufficient to approximate any density function to arbitrary accuracy, given that the mixture coefficient $\alpha_{i}$, the mean and standard deviation of each Gaussian kernel are properly selected \cite{mclachlan1988mixture}.

The output of a MDN is a mixture model parameterized with the mixture coefficient  $\alpha_{i}$, the mean $\mu_{ik}$ and the standard deviation $\sigma_{ik}$. Consequently one can train a MDN to approximate a conditional pdf $p(\mathbf{m}|\mathbf{d})$ for any input data $\mathbf{d}$ by using a set of data-model pairs $\{\mathbf{d}_{i},\mathbf{m}_{i}\}$, where $\mathbf{m}_{i}$ is generated from the prior pdf and $\mathbf{d}_{i}$ is computed from $\mathbf{m}_{i}$ using a known forward function. The network training is performed by minimizing the negative logarithm of the pdf in equation \ref{eq:mixture_model}, which is equivalent to maximizing the likelihood of training data \cite{bishop2006pattern}. After training, the network can be used to estimate the posterior pdf $p(\mathbf{m}|\mathbf{d}_{\mathrm{obs}})$ for any observed data $\mathbf{d}_{\mathrm{obs}}$.

\subsection{Graph mixture density networks}

Graph MDNs combine graph neural networks (GNNs) with mixture density networks, such that the network can output probability distributions for graph data \cite{errica2021graph}. Graphs are a type of data structure which describes a set of objects (nodes) and their relationships (edges), and can be used to provide flexible representation of complex systems. In this study we use graphs to represent travel time data in seismic surface wave tomography, where stations are treated as nodes and edges between stations (nodes) are used to indicate existence of travel time data. The coordinates of stations are set as node features and travel times between stations are set as edge features.  

GNNs are neural networks that manipulate graphs, and can be used within MDNs, called graph MDNs, to predict conditional probability distributions for graph inputs. Graph MDNs have been used in seismic surface wave tomography to predict posterior distribution of phase or group velocity for variable sizes of data across time or frequency \cite{zhang2024rapid}. In this study we apply graph MDNs to surface wave tomographic problems that have different station distributions at sites with various scales, to demonstrate that one can train a graph MDN model to solve many different tomographic problems in practice.  

\begin{figure}[h]
	\centering
	\includegraphics[width=0.9\textwidth]{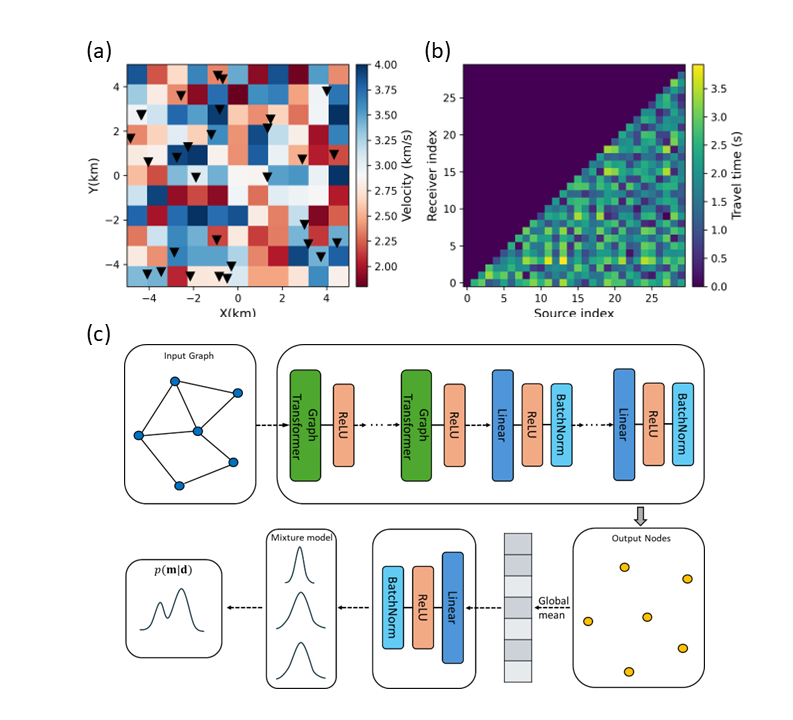}
	\caption{An example of \textbf{(a)} a velocity model and \textbf{(b)} associated travel times randomly selected from the training data. Black triangles denote seismometers whose locations are randomly assigned within the region. \textbf{(c)} The architecture of the graph MDN designed for seismic tomography. The network uses graph transformer layers and graph linear layers to update node features of the input graph, which are then fed into a global mean pooling layer and a linear layer to output a mixture model.} \label{fig_graphMDN}
\end{figure} 

We use a design of graph MDN similar to that in \cite{zhang2024rapid}, which consists of a graph network, a global mean pooling layer, a multilayer perceptron network and a mixture model (Fig \ref{fig_graphMDN}c). The graph network is composed of graph transformer layers and graph linear layers, which update node features of the input graph. Assume $\mathbf{x}_{i}$ as the feature vector of the $i^{th}$ node, and $\mathbf{e}_{ij}$ as the feature vector of the edge between the $i^{th}$ and $j^{th}$ node, a graph transform layer is defined as:
\begin{align}
	\mathbf{x}^{'}_{i} &= \bigg\Vert_{c=1}^{C}\bigg[\mathbf{W}_{1}^{c}\mathbf{x}_{i} + \sum_{j\in\mathcal{N}(i)} \alpha_{ij}^{c}(\mathbf{W}_{2}^{c}\mathbf{x}_{j} + \mathbf{W}_{5}^{c}\mathbf{e}_{ij})\bigg] \nonumber \\
	\alpha_{ij}^{c} &= \mathrm{softmax}\left( \frac{(\mathbf{W}_{3}^{c}\mathbf{x}_{i})^{\mathrm{T}}(\mathbf{W}_{4}^{c}\mathbf{x}_{j}+\mathbf{W}_{5}^{c}\mathbf{e}_{ij})}{\sqrt{d}} \right)
	\label{eq:graph_multhead_transform}
\end{align}
where $\mathbf{x}^{'}_{i}$ is the updated node feature, $\{\mathbf{W}_{m},m=1,2,3,4,5\}$ are trainable weights of the neural network, and $\mathcal{N}(i)$ represents neighbors of the $i^{th}$ node. The symbol $\Vert$ represents concatenation of $C$ vectors (called head attentions), $c$ denotes the $c^{th}$ attention, and $d$ is the dimension of $\mathbf{W}_{3}^{c}\mathbf{x}_{i}$. The updated node features $\mathbf{x}^{'}_{i}$ are then used to calculate the mean feature across nodes, which are fed into a linear layer to estimate the posterior distribution. 


\section{Results}

\subsection{Network training}

We train a graph MDN to estimate posterior pdfs for 2D seismic travel time tomography. The architecture of the graph MDN is shown in Fig. \ref{fig_graphMDN}c. The network consists of 6 graph transformer layers, 4 graph linear layers and a mixture model with 10 Gaussian kernels. The number of channels of the 6 transformer layers is set to 16, 64, 256, 1024, 512 and 512 respectively, and the number of head attentions is set to 4 for the first three layers and 1 for the remaining layers. The number of hidden units for the 4 linear layers is set to 1000, 600, 600 and 1000 respectively. After the graph linear layers, a global mean pooling layer is used to calculate mean features across the nodes, which are fed into a linear layer with 1200 hidden units to output a mixture model.

The velocity model is parameterized using a $11 \times 11$ regular grid with a spacing of 1 km for a $10 \times 10$ km region. At each grid point the prior distribution is set as a Uniform distribution between 1.8 and 4.0 km/s, from which we generate 200,000 velocity structures (Fig. \ref{fig_graphMDN}a). For each structure we deploy 30 seismometers whose locations are randomly assigned within the region, and compute inter-receiver travel times (Fig. \ref{fig_graphMDN}b) using a fast marching method \cite{rawlinson2004multiple}. Gaussian noise with standard deviations of 2 percent of travel times is added to the data. We represent the travel time data using a graph as described above, where receiver coordinates (node features) are normalized using the scale of the region and edge features are set as the path-averaged velocity. By doing this, the trained graph MDNs can be applied to tomographic problems that have different scales compared to the training set, because path-averaged velocity remains the same for a given velocity structure when the scale of the study area and receiver distribution change proportionally.

For network training we use 90 percent of velocity structure and travel time pairs as training data and the remaining 10 percent as test data. The training is performed using the ADAM optimizer \cite{kingma2014adam} for 1,000 iterations with a batch size of 500 and a learning rate of $2 \times 10^{-5}$. To enable variable sizes of input, we randomly drop stations with a ratio of 0.5 and travel times with a ratio of 0.6 at each iteration. The trained network is then used to estimate posterior distributions for newly observed data.

\begin{figure}[h]
	\centering
	\includegraphics[width=1.\textwidth]{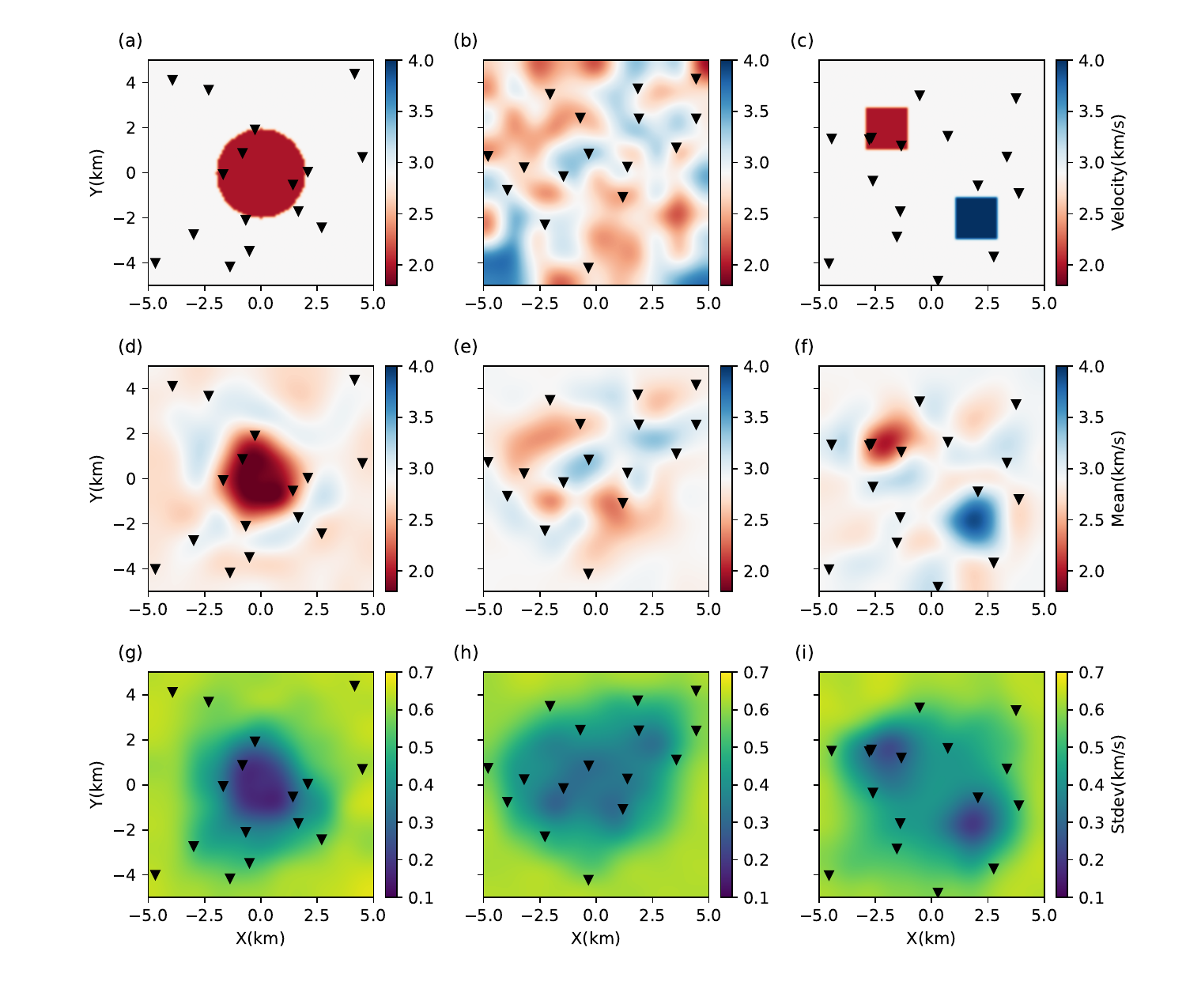}
	\caption{\textbf{(a-c)} The true velocity structure. \textbf{(d-f)} and \textbf{(g-i)} show the mean and standard deviation maps obtained using graph MDN respectively. Black triangles denote locations of seismometers.} \label{figA_synthetic}
\end{figure}

\begin{figure}[h]
	\centering
	\includegraphics[width=1.\textwidth]{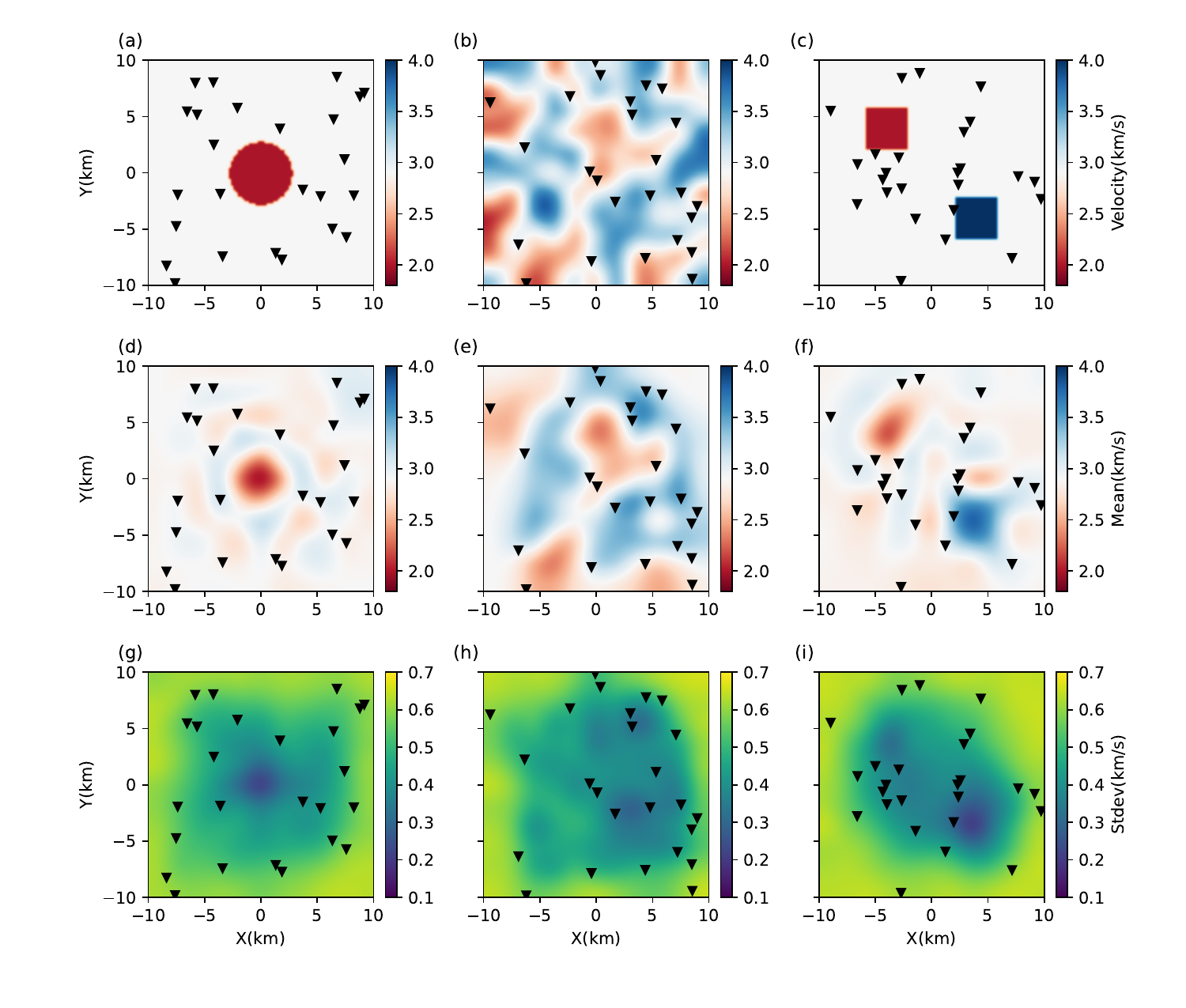}
	\caption{Results obtained using graph MDN for a larger region ($20 \times 20 $km). Keys as in Fig. \ref{figA_synthetic}.} \label{figA_synthetic_scaled}
\end{figure}

\subsection{Synthetic examples}

We first demonstrate the method using a set of synthetic examples. Specifically, we designed three different velocity structures: one has a circle low velocity anomaly within a homogeneous background, one has random velocity structures, and the other one has two square anomalies within a homogeneous background. To demonstrate performance of the method in various acquisition settings, we conducted two sets of tests. In the first set we deployed 16 random seismometers within a same size region as that in the training dataset for each velocity structure, while in the second set we deployed 25 random seismometers within a $20 \times 20$ km region which is larger than that in the training set. For both sets of tests, we simulated inter-receiver travel times on a regular $101 \times 101$ grid using the fast marching method \cite{rawlinson2004multiple}. These travel times are then converted to path-averaged velocity and fed into the trained neural network to estimate posterior pdfs.

The results are shown in Fig. \ref{figA_synthetic} and \ref{figA_synthetic_scaled}. In both sets of tests the mean model shows similar structures to the true model in the area with station coverage, whereas those areas without station distribution show velocity values that are close to the mean of the prior distribution because no ray path travels through these regions. This can also be observed by lower uncertainty in the station covered region and higher uncertainty in the uncovered region as one would expect (Fig. \ref{figA_synthetic}g, \ref{figA_synthetic}h, \ref{figA_synthetic}i and Fig. \ref{figA_synthetic_scaled}g, \ref{figA_synthetic_scaled}h, \ref{figA_synthetic_scaled}i.). The standard deviation maps show lower uncertainties in the region of velocity anomalies, which has also been observed in a range of previous studies \cite{galetti2015uncertainty, zhang2020seismic, zhang2020variational, gebraad2019bayesian, zhang20233} and reflects that these anomalies are well constrained by the data. In addition, the mean model obtained in the second set of tests matches the true model more accurately than those obtained in the first set because there are more stations in the second set of tests. These results therefore demonstrate that graph MDNs can be used to estimate posterior distributions for tomographic problems with different acquisition settings, different number of data and different scales of study region.  

\begin{figure}[h]
	\centering
	\includegraphics[width=1.\textwidth]{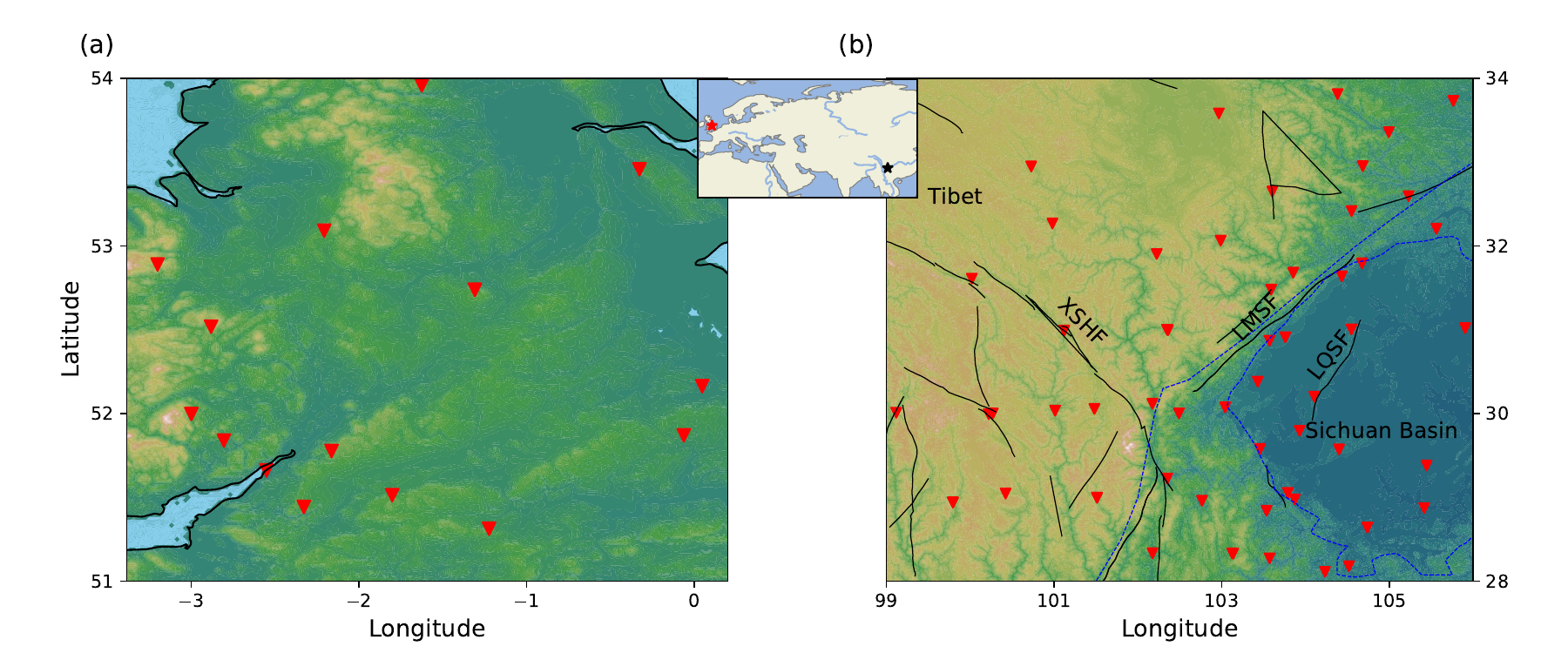}
	\caption{Topography maps and seismometer locations (red triangles) at two study areas in \textbf{(a)} south England and \textbf{(b)} southwest China. Solid black lines show main faults and blue dashed lines indicate tectonic block boundaries. LMSF: Longmen Shan fault; LQSF: Longquan Shan fault; XSHF: Xianshui He fault. Stars in the inset map denote locations of the two area.} \label{fig2}
\end{figure}

\subsection{Application to South England}

To further demonstrate the method, we apply the trained graph MDN to two real datasets. For the first test, we use the graph MDN to estimate posterior distributions of group velocity at a set of periods for Love waves extract from ambient noise data  in South England (Fig. \ref{fig2}a). The ambient noise data were recorded by 15 seismometers in 2006 - 2007 and 2010, and processed to produce path-averaged group velocities between each station pair (details can be found in \cite{galetti2017transdimensional}). These group velocity measurements are then fed into the graph MDN to predict posterior distributions for each period. Note that the scale of the problem and the station distributions are different with those in the training data.

\begin{figure}[h]
	\centering
	\includegraphics[width=1.\textwidth]{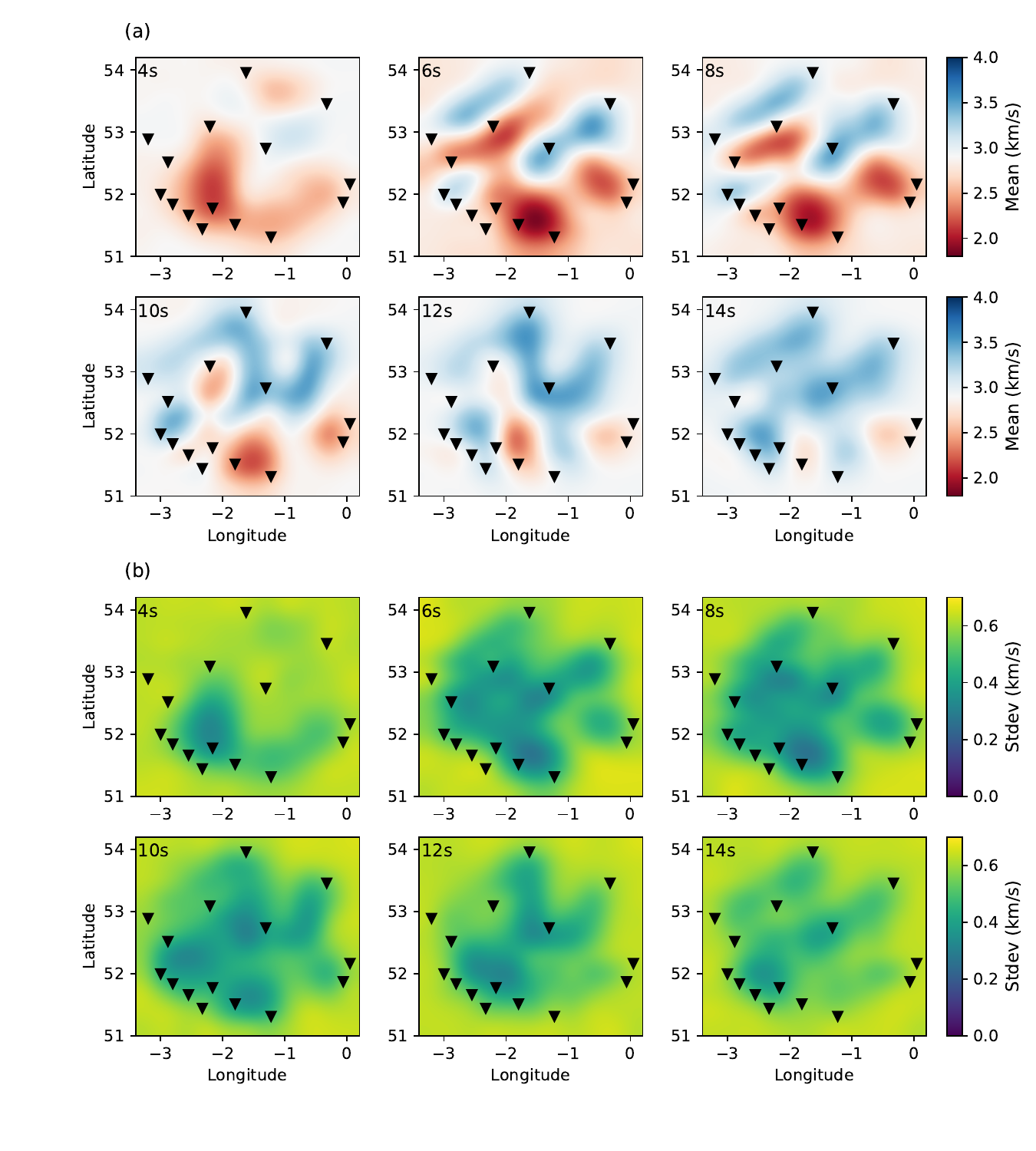}
	\caption{The \textbf{(a)} mean and \textbf{(b)} standard deviation maps obtained using graph MDN at periods of 4 s, 6 s, 8 s, 10 s, 12 s and 14 s in south England.} \label{fig3}
\end{figure}

Fig. \ref{fig3} shows the mean and standard deviation maps obtained at different periods. Overall the structures of the mean velocity maps reflect various geology features at shallow depth in the region, with sedimentary rocks being displayed as low velocity while igneous and metamorphic complexes are shown as high velocity anomalies \cite{galetti2017transdimensional}. For example, there are low velocity anomalies in the south of the region (around -1.8\textdegree E, 51.8\textdegree N), in the west-central area (-2\textdegree E, 53\textdegree N), and in the southeastern region (-0.2\textdegree E, 52.1\textdegree N), which are associated with the Midland Platform, the Cheshire Basin and the Anglian Basin, respectively \cite{galetti2017transdimensional}. The anomalies in the south and southeast are visible across all the periods, whereas that in the west-central area disappears at 14 s. This likely reflects that the sendimentary thickness of the Cheshire Basin is smaller than other basins. Similarly there is a high velocity anomaly in the northwest (around -2.5\textdegree E, 53.5\textdegree N) following approximately a northeast-southwest trend associated with the limestone of the Pennines. In the northeast (around -1.5\textdegree E, 53\textdegree N) a high velocity anomaly with similar trend can also be observed, which may reflect the granitic batholiths and dykes underneath the area, or be interpreted as the evidence of Proterozoic basement in an area of thin sedimentary basin \cite{galetti2017transdimensional}. Note that these low and high velocity anomalies have smaller values at 4 s period compared to those at other periods. This is likely caused by the lower ray coverage at 4 s period, which can also be observed by the higher uncertainty in the standard deviation map.

Overall the standard deviation maps show similar structures as the mean velocity. For example, at locations of these low and high velocity anomalies the results show lower uncertainty, which indicates that these anomalies are well constrained by the data. Similar phenomena have also been observed by a various of previous studies \cite{zhang2020seismic, gebraad2019bayesian, zhang2020variational}. By contrast, in areas outside the receiver array the results show higher uncertainty because few ray paths go through these regions.  

To further analyse the results, we compare them with those obtained using Markov chain Monte Carlo (McMC). For each period we use a standard adaptive Metropolis-Hastings algorithm \cite{haario2001adaptive, salvatier2016probabilistic} to generate posterior samples with a total of 6 chains, each of which contains 1,000,000 samples including a burn-in period of 500,000; the burn-in samples are discarded and every 10th of the remaining 500,000 samples are retained to calculate statistics of the posterior pdf.

\begin{figure}[h]
	\centering
	\includegraphics[width=1.\textwidth]{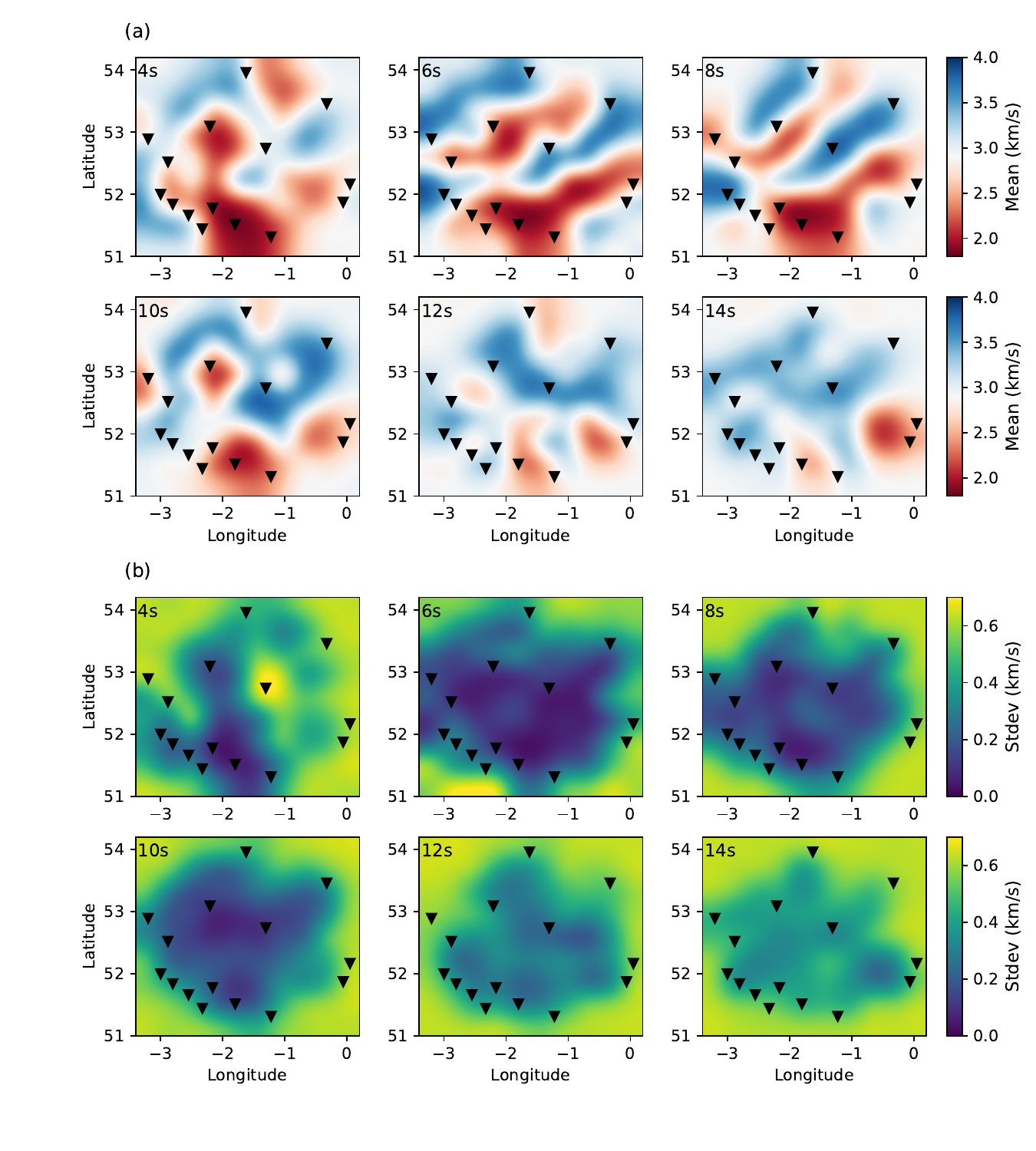}
	\caption{The \textbf{(a)} mean and \textbf{(b)} standard deviation maps obtained using McMC at periods of 5 s, 10 s, 15 s, 20 s, 25 s and 30 s in south England.} \label{figA_uk_mcmc}
\end{figure}

Fig. \ref{figA_uk_mcmc} shows the results obtained using McMC. Overall the mean and standard deviation maps show similar features to those obtained using graph MDN, which demonstrates that graph MDN can provide accurate estimates of posterior distributions even the network is trained on a much smaller scale (Fig. \ref{fig_graphMDN}a). However, there exists small details which are differ in the two results. For example, the shape of the low and high velocity anomalies is slightly different. In addition, the results obtained using McMC show higher magnitude of velocity variations and lower uncertainty. These inconsistencies are likely caused by insufficient training of the network because of difficulty in training MDNs in a high dimensionality \cite{hjorth1999regularisation, makansi2019overcoming, zhang2021bayesianb}. Nevertheless, the two results are broadly consistent.

\subsection{Application to Southwest China}

As a second example we apply the graph MDN to Rayleigh wave phase velocity measurements extracted from ambient noise data that were recorded by 57 seismometers in Southwest China (Fig. \ref{fig2}b). Details of these phase velocity measurements can be found in \cite{liu2021community}. We fed these measurements to the trained graph MDN to estimate posterior distributions of phase velocity at periods from 5 s to 30 s. This example demonstrates an application where more stations are available than those in the training data.

\begin{figure}[h]
	\centering
	\includegraphics[width=1.\textwidth]{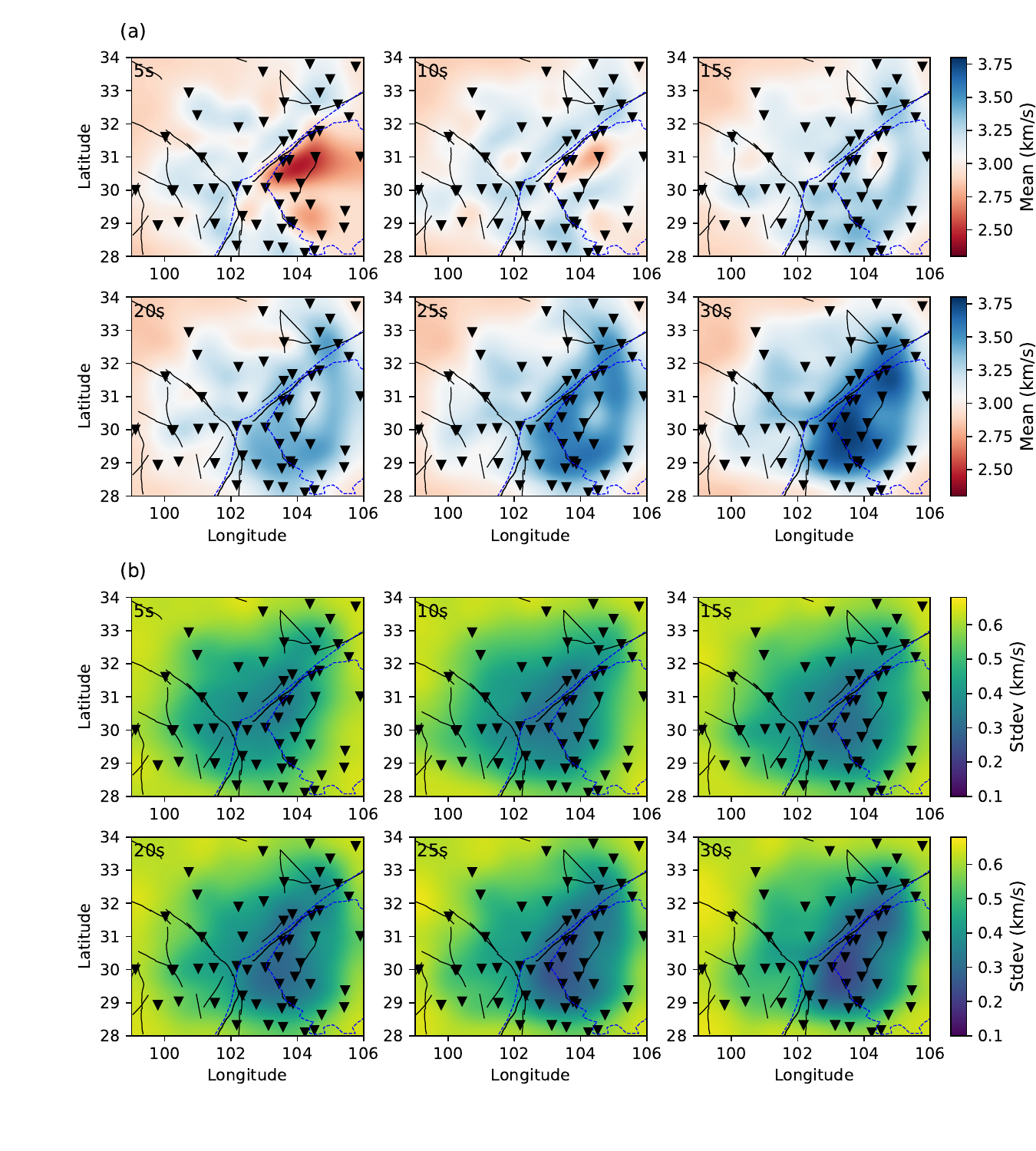}
	\caption{The \textbf{(a)} mean and \textbf{(b)} standard deviation maps obtained using graph MDN at periods of 5 s, 10 s, 15 s, 20 s, 25 s and 30 s in southwest of China.} \label{fig4}
\end{figure}

Fig. \ref{fig4} shows the mean and standard deviation maps obtained at different periods. Overall the mean maps show similar features to those found by previous studies \cite{bao2015two, yang2020new, liu2021community, liu2023high}. For example, there are clear velocity variations across the Longmen Shan fault (LMSF). At short periods (5 - 10 s) the results show lower velocity in the east of the fault which is associated with the Sichuan basin, whereas in the west the Tibet shows higher velocity. By contrast, at longer periods (15 -30 s) the Sichuan basin displays higher velocity than that in the Tibet. This likely reflects that at those periods the phase velocity has been affected by velocity of upper mantle in the Sichuan basin because of the smaller crustal thickness compared to that of the Tibet. Within the Sichuan Basin there is a relatively higher velocity anomaly at short periods in the central related to the uplift zone in the region. In addition, a relatively lower velocity anomaly can be observed across all the periods at the location of the Longquan Shan fault (LQSF), which may be related to the activity of the fault.

Similarly to the previous example, the standard deviation maps show higher uncertainty outside the receiver array because of lower ray coverage. Within the array the uncertainty is generally lower along the LMSF and in the southwest of the Sichuan basin, which is likely because of the denser station distribution in these regions. Note that the uncertainty maps show smoother structures compared to those obtained in the previous example. This is probably caused by the denser ray coverage, which makes the region roughly equally constrained by the data and hence leads to smoother structures.

Due to the larger data set we compare the results with those obtained using the Stein variational gradient descent (SVGD) method, which has been demonstrated to be more efficient than McMC methods \cite{liu2016stein, zhang2020seismic}. We generate 500 particles from the prior distribution, and update them using SVGD for 1,000 iterations. These final set of particles are then used to calculate statistics of posterior distributions.  

\begin{figure}[h]
	\centering
	\includegraphics[width=1.\textwidth]{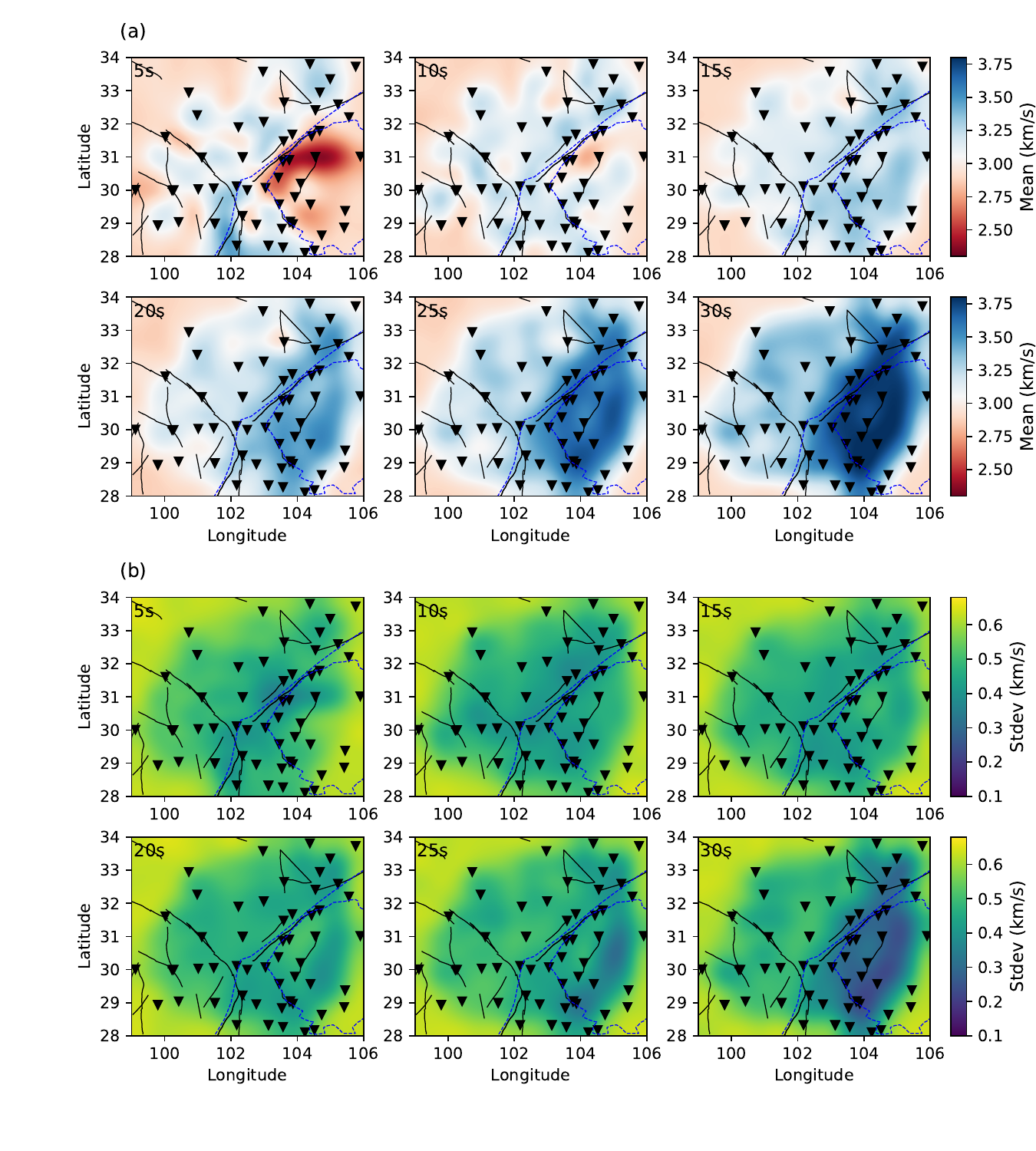}
	\caption{The \textbf{(a)} mean and \textbf{(b)} standard deviation maps obtained using SVGD at periods of 5 s, 10 s, 15 s, 20 s, 25 s and 30 s in southwest of China.} \label{figA_Tibet_svgd}
\end{figure}

Overall the results show similar mean and standard deviation maps to those obtained using graph MDN (Fig. \ref{figA_Tibet_svgd}). For example, on the two sides of LMSF there are clear low and high velocity contrasts associated with the Sichuan Basin and Tibet respectively. The low velocity anomaly associated with the LQSF can also be observed. This again demonstrates that graph MDNs can provide accurate estimates of posterior distributions for various problems. Similarly as above there are small details that differ in the two results, which may be caused by insufficient training of the graph MDN, or because SVGD has not converged sufficiently.
 
\section{Discussion}

Graph MDNs can produce estimates of posterior distributions very efficiently. For example, the above tests require 0.45 hours to generate the training data using 6 Intel(R) Xeon(R) Gold 6258R CPU cores, and 59.3 hours to train the neural network using a NVIDIA GeForce RTX 4090 GPU card. Once trained, the network produces estimates of posterior distribution in just 0.6 seconds for each period. By contrast, to obtain the above results it took about 16 hours for McMC and 3.9 hours for SVGD for each period using 6 CPU cores.

We have shown that graph MDNs can be used to solve tomographic problems with various station distributions and scales at different sites. This therefore demonstrates the possibility to train one single neural network model to provide solutions for various tomographic problems. Note that we used a fixed regular grid to parameterize the velocity model, which may become inappropriate if a finer or coarser grid is required. In this case, one can train a set of network models using various grids to satisfy different spatial resolution requirements. In addition, one may add velocity as extra node features and use graphs to parameterize the velocity model to provide more flexible representation of the subsurface.

Although we only applied graph MDNs to 2D tomography, the method can be extended straightforwardly to other tomographic problems. For example, one can use graph MDNs to solve 3D body wave tomographic problems by treating both sources and receivers as graph nodes, or to solve full waveform inversion problems by setting waveforms as node features. However, we note that this may require much larger networks to be trained.

\section{Conclusion}

In this study we trained a graph MDN for 2D tomography using simulated data with random velocity and station distributions, and applied the network to both synthetic and real datasets which have different scales and station distributions compared to the training data. The results show that the network can  provide comparable estimates of posterior distributions compared to those obtained using Monte Carlo and SVGD methods in less than a second. This therefore demonstrates the possibility to solve efficiently various tomographic problems using one or a set of deep learning models.  

\section*{Acknowledgments}
The authors thank the National Science and Technology Major Project of China (2024ZD1000403), National Natural Science Foundation of China (42204055 and U23B20160) and the Fundamental Research Funds for the Central Universities of China University of Geosciences in Beijing for supporting this research.

\bibliographystyle{plainnat}
\bibliography{graph}

\label{lastpage}

\end{document}